\documentclass[reprint,prb,showpacs,superscriptaddress,aps]{revtex4-1}

\usepackage{graphicx}
\DeclareGraphicsExtensions{.png,.jpg,.eps}

\usepackage{float}
\usepackage{xcolor}
\usepackage{amsmath}
\usepackage{caption}
\usepackage{subcaption}
\usepackage{float}

\usepackage{wrapfig}

\begin{document}

\title[Sample title]{Oxygen vacancies in strained SrTiO$_{3}$ thin films: formation enthalpy and manipulation.}

\author{Lucia Iglesias}
\affiliation{Centro de Investigacion en Quimica Bioloxica e Materiais Moleculares (CIQUS) and Departamento de Quimica-Fisica, Universidade de Santiago de Compostela, 15782-Santiago de Compostela, Spain.}
\author{Alexandros Sarantopoulos}
\affiliation{Centro de Investigacion en Quimica Bioloxica e Materiais Moleculares (CIQUS) and Departamento de Quimica-Fisica, Universidade de Santiago de Compostela, 15782-Santiago de Compostela, Spain.}
\author{C. Magen}
\affiliation{Laboratorio de Microscopias Avanzadas (LMA), Instituto de Nanociencia de Aragon (INA) - ARAID, 50018 Zaragoza, Spain}
\affiliation{Departamento de Fisica de la Materia Condensada, Universidad de Zaragoza, 50018 Zaragoza, Spain}
\author{F. Rivadulla}
\email{f.rivadulla@usc.es}
\affiliation{Centro de Investigacion en Quimica Bioloxica e Materiais Moleculares (CIQUS) and Departamento de Quimica-Fisica, Universidade de Santiago de Compostela, 15782-Santiago de Compostela, Spain.}

\date{\today}

\begin{abstract}
We report the enthalpy of oxygen vacancy formation in thin films of electron-doped SrTiO$_{3}$, under different degrees of epitaxial stress. We demonstrate that both compressive and tensile strain decrease this energy at a very similar rate, and promote the formation of stable doubly ionized oxygen vacancies. Moreover, we also show that unintentional cationic vacancies introduced under typical growth conditions, produce a characteristic rotation pattern of TiO$_6$ octahedra. The local concentration of oxygen vacancies can be modulated by an electric field with an AFM tip, changing not only the local electrical potential, but also producing a non-volatile mechanical response whose sign (up/down) can be reversed by the electric field.

\end{abstract}


\maketitle

\section{Introduction}

The flexibility of oxoperovskites for accepting a large number of cationic dopants is extremely useful for the systematic study of the correlations between crystalline structure, electronic structure, and functionality.  Their high chemical stability also makes them well suited materials for growth in single-crystalline and thin film form. Combining these characteristics with the mixed valence states of 3$d$ transition metal ions, results in some of the most remarkable properties and functionalities found in  solids.\cite{GoodenoughMCB,TsudaSSS}
A good example of this richness is the case of SrTiO$_3$ (STO), a diamagnetic quantum paraelectric insulator \cite{Montu}, in which cationic vacancies induce polar regions of nanometer size stabilized by stress.  \cite{CBE_Science,PRLCBE} 
 Oxygen vacancies, (V$_{O}$), are also readily formed in this material; due to their donor-character and the very large electron mobilities characteristic of delta-doped STO, even the slightest concentration of vacancies produces a measurable electrical conductivity.\cite{Son2010,Santander-Syro2010,Meevasana2011,Reagor2005}
 Actually, reduced STO was the first oxide reported to show superconductivity, with a maximum T$_C$$\approx$0.3 K for electron densities of $n$$\approx$10$^{20}$ cm$^{-3}$.\cite{Schooley,Schooley2}

The unintended presence of oxygen vacancies could be particularly important in thin films of STO: first, typical deposition conditions imply high temperature and low oxygen pressure, which favors the reduction of STO. Second, donation of the extra charge of ionized oxygen vacancies to non-bonding orbitals of the perovskite results in expansion and local TiO$_6$-octahedral  rotation.\cite{PetrieADFM, GazquezSrCoO3}
Ab-initio calculations actually support a lowering of the V$_{O}$ formation energy under tensile strain for SrTiO$_3$ \cite{ChoiNL} and other perovskite oxides, like CaMnO$_3$,\cite{SpaldinPRBCaMnO3,SpaldinNL2017} LaAlO$_3$,\cite{PhysRevB.88.174101} or SrCoO$_{3-\delta}$. \cite{PetrieADFM}
The same chemical expansion mechanism should in principle disfavor anionic vacancy formation under compressive stress. In this case several calculations have been published in the last few years, but the results are diverse for different oxides: in CaMnO$_3$, the calculated effect of a moderate compressive stress is almost negligible,\cite{SpaldinPRBCaMnO3} while it results in a considerable increase of V$_{O}$ formation energy in LaAlO$_3$,\cite{PhysRevB.88.174101} SrCoO$_{3-\delta}$ \cite{PetrieADFM} and ferroelectric BaTiO$_3$.\cite{YangJAP1} 
For the case of SrTiO$_3$, however, different authors suggest that both compressive and tensile strain can stabilize V$_{O}$. \cite{HamadanyJAP,ChoiNL,Zhang} 

It is vital the inclusion of the precise pattern of TiO$_6$ octahedral rotation and the actual charge of the vacancies in the calculations, to obtain reliable results. Furthermore, the presence of cationic vacancies can have a strong influence and should be taken into account, although rarely are considered jointly with V$_O$ in the calculations.\cite{RondinelliAM} 
 
Given the relevance of oxygen vacancy formation in the functional properties (e.g. resistive switching, improved catalytic properties, etc) of STO thin films and other oxides,\cite{PetrieJACS,TahiniJACS} it is very important to experimentally measure the effect of tensile and compressive stress on the formation energy of oxygen vacancies in STO thin films.
 
Here, we report a thermodynamic study of the process of oxygen vacancy formation in thin films of electron-doped STO under different degrees of epitaxial strain. Our main finding is the experimental demonstration of a significant reduction of the enthalpy of oxygen vacancy formation, for both compressive and tensile stress. This points to a common mechanism irrespective to the sign of epitaxial strain, most probably due to a reduction of the energy gap. Furthermore, we show that even though the actual TiO$_6$-octahedral rotation pattern observed in the film is due to the presence of cationic vacancies, they are cooperative throughout the whole film, and not just around point defects. Finally, we demonstrate that oxygen vacancy manipulation by an electric field can lead to non-volatile changes in the local volume. This offers the possibility of mechanical control of the local electrical properties.


\section{Experimental Details}

Oxygen vacancies are characteristic e-donor defects and therefore their concentration can be determined by Hall effect experiments. In order to detect small concentrations of oxygen vacancies created at relatively high deposition oxygen pressure and high temperature, any source of intrinsic acceptor impurities should be previously neutralized. It is known that nominally undoped SrTiO$_3$ single crystals contain small concentrations of acceptor impurities.\cite{DeSouzaPRB} These occur mainly in the form of ionized Sr vacancies (V$_{Sr}$), which trap part of the electrons donated by the V$_O$ to the conduction band of the oxide. The presence of V$_{Sr}$ is even more important in thin films, particularly in those deposited by Pulsed Laser Deposition (PLD). Achieving the correct stoichiometry in this case requires a very fine tuning of the laser energy, substrate temperature, and background oxygen pressure during deposition. Furthermore, our acceptor-doped films show an accumulation of oxygen vacancies at the surface and a complex dependence of the conductivity with oxygen pressure.\cite{DeSouzaPRB2014}
Therefore, in order to minimize the effect of extrinsic acceptor impurities over the charge donated by V$_O$, we have chosen Nb-doped  SrTiO$_3$ for this study. Around 2\% Nb results in a density of free electrons which compensates the presence of intrinsic acceptor impurities, as measured by Hall effect.\cite{Sarantopoulos} As a consequence, even small amounts of V$_O$ produce measurable changes in the conductivity and Hall effect. On the other hand, we have kept the concentration of Nb small enough for the oxygen vacancies to produce an observable effect at any pressure.

Following this approach, Nb:STO thin films  were grown by Pulsed Laser Deposition (KrF, $\lambda$=248 nm) on different substrates (see supporting information for a complete list of the substrates and orientations used in this work), with lattice parameters ranging from pseudocubic a$_s$=3.791 $\AA$ of LaAlO$_{3}$ to a$_s$=3.970 $\AA$ of GdScO$_{3}$, changing the epitaxial stress in the interval s(\%)=[-2.91, +1.66] (s(\%)=$\frac{a_s-3.905}{3.905}$$\times$100; 3.905 \AA refer to the lattice parameter of bulk STO).
 The deposition conditions were previously optimized in order to minimize the presence of cationic vacancies in the as-deposited films:\cite{Sarantopoulos} laser fluence $\approx{}$0.9 J/cm$^{2}$, repetition rate 4 Hz, 800\textordmasculine{}C under oxygen pressure of P(O$_{2}$)= 100 mTorr.  The films were cooled down to room temperature at 5\textordmasculine{}C/min under the same atmosphere. The nominal Nb concentration, $\approx$2$\%$, and the thickness $\approx$20$\pm$2 nm, were kept constant for all films studied in this work. 

A controlled amount of oxygen vacancies was introduced in the films by post-deposition annealing at different temperature (either 800\textordmasculine{}C or 600\textordmasculine{}C ) and oxygen pressure, P(O$_{2}$), according to the following process:

\begin{equation}\label{eq1}
SrTiO_3\ (s)\rightleftharpoons{}SrTiO_{3-x}(s)+\ \frac{x}{2}O_2(g)\uparrow{}
\end{equation}

Samples were left to reach equilibrium with the background oxygen pressure for 2 hours, and then rapidly quenched to room temperature at a cooling rate $\approx$100 \textordmasculine{}C/min, to keep the concentration of vacancies unchanged during cooling. The [V$_O$] was quantified immediately after this by Hall effect measurements, using the Van der Pauw configuration.

The local distribution of oxygen vacancies was studied using Electrostatic Force Microscopy (EFM mode), in an NX-10 Atomic Force Microscope (Park Systems).


\section{Results and Discussion}

\begin{figure}[h]
\begin{center}
\includegraphics[width=200pt]{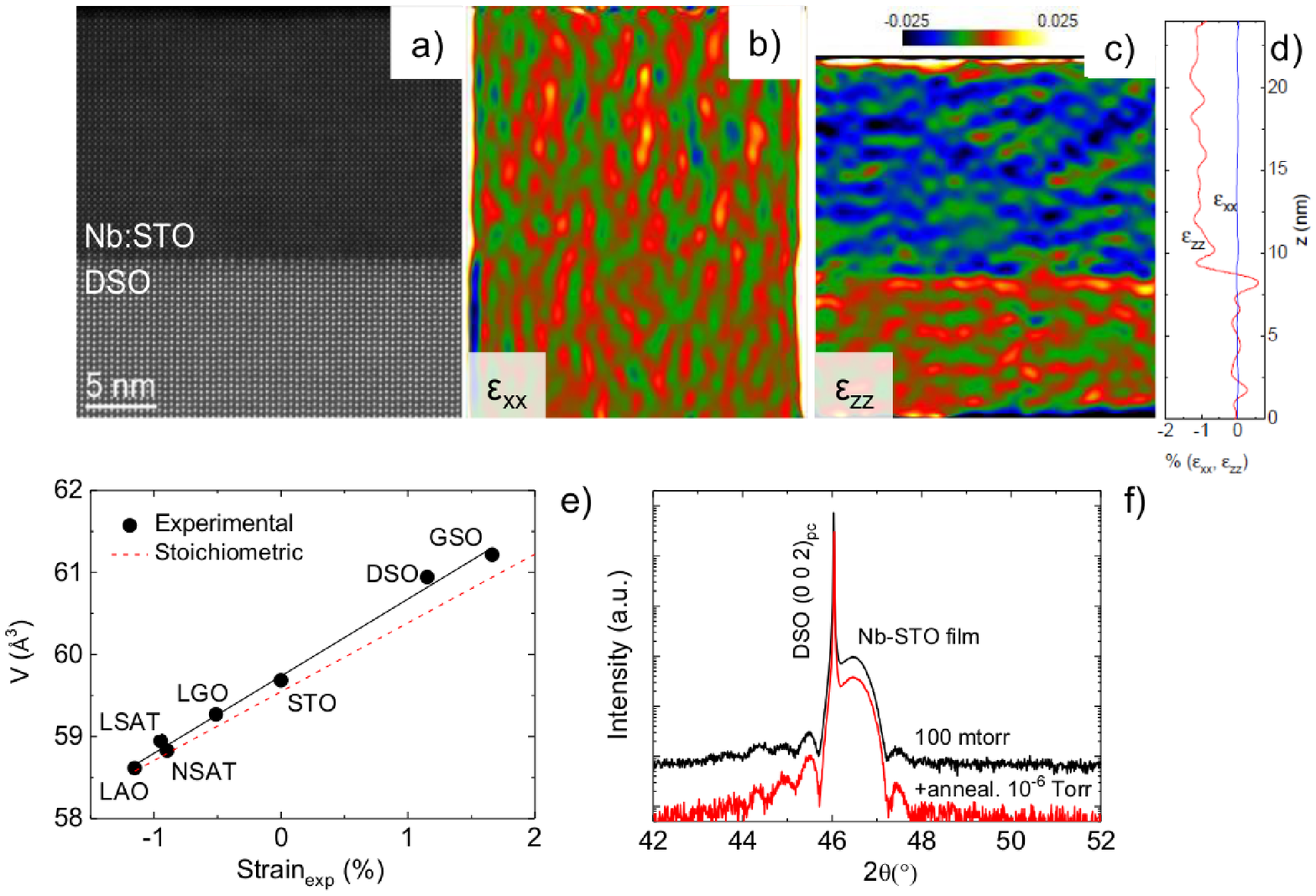}
\caption{HAADF-STEM image of as-grown Nb:STO thin films deposited on DSO (a). The corresponding GPA analysis with the in-plane $\varepsilon_{xx}$ and out of plane $\varepsilon_{zz}$ contraction with respect to the substrate are shown in b) and c), along with the corresponding line profile in d). e)  Evolution of the unit cell volume with experimental strain for all the samples used in this study. The dashed line indicates the theoretical volume of the stoichiometric unit cell with a Poisson ratio of $\nu{}$=0.23. The solid line is a guide to the eye. f) XRD $\theta{}$-2$\theta{}$ scan of the sample as-grown at 100 mTorr and after the annealing process at 1x10$^{-6}$ Torr, showing the negligible effect on the lattice parameters.}\label{fig:1}
\end{center}
\end{figure}

High-resolution reciprocal space maps (RSM, see supporting information Figure S1) indicate a good epitaxial matching to the substrates for all films, except for these grown on LAO. This is the substrate with the smallest lattice parameter used in this work, and the films are already partially relaxed.
As an example of the crystalline quality of the films studied, the Geometrical Phase Analysis (GPA) of High Angle Annular Dark Field-Scanning Transmission Electron Microscopy (HAADF-STEM) images is shown in Figures \ref{fig:1}a) to d). The GPA analysis also confirms that the out of plane  positive (negative) distortion ($\epsilon{}$$_{zz}$) in response to the compressive (tensile) in-plane strain ($\epsilon{}$$_{xx}$) is homogeneous throughout the films, and in complete agreement with the RSM data. For example, for the STO deposited on DSO the c-axis parameter of the film contracts $\approx$1.1\% with respect to the substrate (see Figure \ref{fig:1}d) and Figures S1 and S2 in the supporting information).

The change of volume with strain was calculated from the X-ray data, and is shown in Figure \ref{fig:1}e) for the as-grown Nb:STO films. This is compared to the theoretical volume assuming the Poisson ratio $\nu$ = 0.23 characteristic of bulk SrTiO$_{3}$.\cite{Ledbetter} Introducing oxygen vacancies by post-annealing has a negligible effect in the volume of the lattice (Figure \ref{fig:1}f), at least for the amount of vacancies explored in this work.\cite{Onishi} There is a clear and progressive deviation of the theoretical prediction for tensile stressed samples. From this point of view, it is particularly interesting the elongation of the out of plane lattice parameter by  $\approx$1$\%$, for the samples deposited on STO substrates, despite the nule epitaxial stress induced by the substrate in these films. The deviation from the value of $\nu$ = 0.23 points towards slight variations of the stoichiometry, namely V$_{Sr}$.\cite{Sarantopoulos}

\begin{figure}[h]
\begin{center}
\includegraphics[width=200pt]{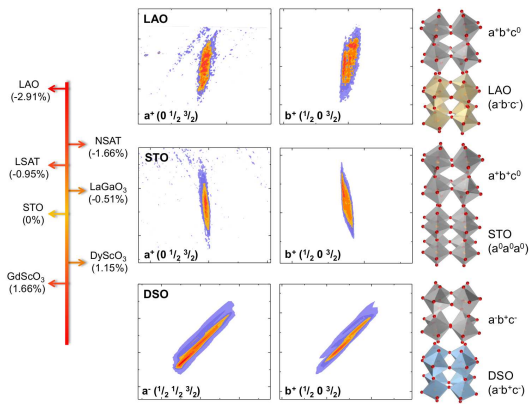}
\caption{High resolution reciprocal space maps around half-order reflections for a Nb:STO thin film deposited on different substrates. The figure shows the rotation pattern characteristic of a$^+$b$^+$c$^0$, found for samples grown on STO and under compressive stress. The percentages of stress induced by the different substrates on the Nb:STO films are indicated along the vertical line on the left of the figure, while the different rotation groups of the film and substrates are shown in the pictures on the right.}\label{fig:2}
\end{center}
\end{figure}

The effect of cationic vacancies on the structural properties of the films was carefully investigated through the presence of different X-ray half-order reflections.
These experiments (see Figure \ref{fig:2}) confirmed a complex rotation pattern of the TiO$_6$ octahedra of the films. In the case of samples deposited on STO, the half-order reflections are characteristic of either an a$^+$b$^+$c$^0$ or an a$^+$a$^+$c$^0$ tilt system in the Glazer notation.\cite{Glazer} These are consistent with a tetragonal or orthorhombic space group, different from the a$^0$b$^0$c$^0$ of cubic STO.  Therefore, given the negligible lattice mismatch between the Nb:STO film and the STO substrate, this result reflects the effect of V$_{Sr}$ introduced during growth over the lattice symmetry of the film. Moreover, the clear observation of the X-ray half-order reflection implies that the tilting of the TiO$_6$ octahedra occurs along the whole sample, not just around a local vacancy.

The tolerance factor, $t$, quantifies the mismatch of the equilibrium A-O/B-O bond-lengths in ABO$_3$ perovskites. For cubic STO $t$=1, also being true for 2$\%$-Nb:STO. However, a progressive reduction of the ionic radius at the A-site is expected to induce a cooperative rotation of the BO$_6$ octahedra, which reduces the symmetry to tetragonal ($I4/mcm$, rotation along [001] axis), rhombohedral ($R-3c$, rotation along [111] axis), and orthorhombic ($Pbnm$ or $Pnma$, rotation along [110] axis) as t decreases.\cite{JBG1}
The presence of V$_{Sr}$ introduces a distortion similar to an average reduction of the ionic radius of the A-cation. This decreases the tolerance factor and puts the Sr-O (Ti-O) bonds under tensile (compressive) stress. The amount of compression of the Sr-O bonds is usually larger than the one of the Ti-O bonds, resulting in dt/dP$<$0, and therefore compressive stress goes in the same direction as the reduction of the ionic radius.
Consistent with this hypothesis, the same rotation pattern was observed for all the films grown under compressive stress, either on cubic LSAT or orthorhombic LAO, in spite of the trigonal $R-3c$ symmetry (a$^-$a$^-$a$^-$) of the latter (see Figure \ref{fig:2}).

 On the other hand, tensile stress has the opposite effect. For the films deposited under tensile stress only half-order reflections corresponding to the a$^-$b$^+$c$^-$ rotations of the substrate (orthorhombic $Pnma$) were observed. The positive stress dominates over the effect of V$_{Sr}$, imposing on the films the rotation pattern of the crystal underneath.

Moreover, we confirmed experimentally that V$_O$ introduced by post-deposition annealing do not affect the rotation pattern of the TiO$_6$ octahedra.\cite{Zhang,PhysRevB.88.174101} Only cationic vacancies determine the octahedral rotation pattern in unstrained and compressed STO films.

These results demonstrate that different octahedral rotation patterns are induced in thin films prepared under similar conditions (identical composition), but subject to different degrees (and sign) of epitaxial stress (i.e. deposited on different substrates). This could play a double role: i) changing the formation enthalpy of a vacancy, given the change in local distances and coordination; and ii) influencing the diffusion coefficient of the oxygen vacancies along the different directions of the perovksite structure.

\begin{figure}[h]
\begin{center}
\includegraphics[width=200pt]{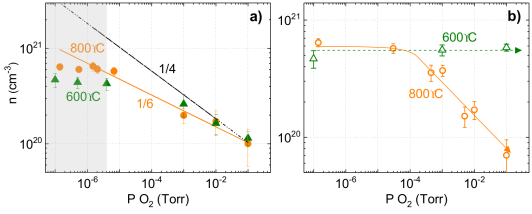}
\caption{a) Variation of the carrier concentration of Nb:STO@LSAT with the oxygen pressure, for samples annealed at 800\textordmasculine{}C (circles) and 600\textordmasculine{}C (triangles). The solid line represents the expected behavior according to equation \ref{eqnvsP}, for doubly ionized oxygen vacancies. b) Re-oxygenation process, annealing the reduced samples at higher oxygen pressures, at 800\textordmasculine{}C (squares) and 600\textordmasculine{}C (triangles). Lines in b) are guides to the eye.}\label{fig:3}
\end{center}
\end{figure}

The formation of oxygen vacancies in SrTiO$_3$ thin films may occur as a Schottky defect to compensate acceptor vacancies Sr$^{2+}$ formed during deposition at low oxygen pressure. Alternatively, V$_O$ can be introduced in stoichiometric films by post-annealing in a reducing atmosphere. 
We introduced a controlled amount of oxygen vacancies in our films by post annealing at high temperature, either at 800ºC or 600ºC, under different oxygen pressures. 
Although V$_O$ in electron-doped STO are not stoichiometric vacancies (cation+anion), a simple mass action law for (\ref{eq1}) is applicable as long as the concentration of vacancies is not too large, so the crystal can be still considered nearly stoichiometric. In this case,  we can formulate the following equilibrium equation for the material annealed at low oxygen pressures:

\begin{equation}\label{eq2}
O^{2-}\rightleftharpoons\frac{1}{2}O_2+V_O+2e
\end{equation}

where O$^{2-}$ represents the oxygen ions at their corresponding STO lattice sites. These can be included in an effective equilibrium constant $K$:

\begin{equation}\label{reaction1}
K=\left[V_O\right]n_e^2P_{O_2}^{1/2}
\end{equation}

If the effect of Sr$^{2+}$-acceptor vacancies is compensated by Nb$^{5+}$-donors originally present in the sample, then the reduction reaction (\ref{eq2}) will produce a change in the carrier density that can be measured by Hall effect experiments. If  n$_e$=2[V$_O$], i.e. each O$^{2-}$  vacancy donates two electrons to the conduction band, substituting this result in the equation for the equilibrium constant, we obtain:

\begin{equation}\label{eqnvsP}
log{n_e}\propto-\frac{1}{6}\log{P_{O_2}}
\end{equation}

An extrinsic source of oxygen vacancies, for example due to acceptor impurities or an excess of TiO$_2$, will render the concentration of vacancies mostly independent of the oxygen pressure; the reduction reaction (\ref{eq2}) will not be the main source determining the concentration of vacancies in this case. According to equation (\ref{reaction1}) the slope of the $n$ vs PO$_2$ plot (equation \ref{eqnvsP}) will change from 1/6 to 1/4 .

The experimental results are shown in Figure \ref{fig:3}a). The carrier density, determined by Hall effect measurements, increases progressively after high temperature-low pressure annealing, up to a saturation value of $n$ $\approx $6-7$\times$10$^{20}$cm$^{-3}$ at $\approx$ 10$^{-5}$Torr of oxygen background atmosphere. The increment of $n$$_e$ as oxygen pressure decreases is in good agreement with the expectations for doubly ionized vacancies (equation \ref{eqnvsP}), confirming the validity of the previous hypothesis. 
On the other hand, while re-annealing the previously deoxygenated samples at higher oxygen pressures and 800\textordmasculine{}C completely recovers the initial state, this is not the case when the samples are annealed at 600\textordmasculine{}C, or lower temperatures (see Figure \ref{fig:3}b). Temperature is a critical factor in the creation/annihilation of oxygen vacancies in Nb:STO, and should be taken into account in the synthesis and post-annealing protocols of fully oxygen-stoichiometric samples. 

\begin{figure}[h]
\begin{center}
\includegraphics[width=200pt]{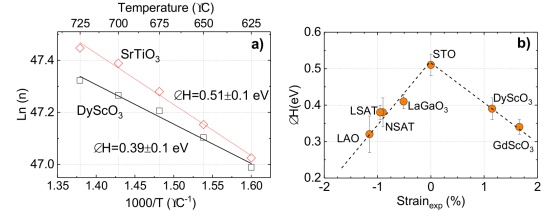}
\caption{a) Carrier concentration as a function of the annealing temperature at 1 mTorr oxygen pressure for the samples grown on DSO and STO substrates. b) Enthalpy of formation of oxygen vacancies as a function of the epitaxial strain. Dashed lines are guides to the eye.}\label{fig:4}
\end{center}
\end{figure}

In addition, the thermodynamic equilibrium constant entering the mass action law is related to the Gibbs free energy of the crystal, and therefore to the enthalpy and entropy of oxygen vacancies formation . Assuming that the change in entropy is only due to the increasing possibilities of configurational arrangements, the equilibrium constant is:

\[
K=K_0e^{-\Delta H/k_BT}
\]
which, through equation (\ref{eqnvsP}) can be related to the temperature dependence of the carrier density through:

\[
\log{n}=K_0'-\frac{1}{9}\log{P_{O_2}}-\frac{\Delta{}H}{3\kappa{}T}
\]
and, at constant pressure yields:

\begin{equation}\label{eqDH}
\log{n}=K_0''-\frac{\Delta{}H}{3\kappa{}T}
\end{equation}

According to the equation (\ref{eqDH}), the formation enthalpy of an oxygen vacancy can be obtained from the slope of an isobaric plot of log $n$$_e$ versus the inverse of the annealing temperature. This is shown in Figure \ref{fig:4} a) for two films, grown respectively on STO (s(\%)=0.0) and DSO (s(\%)=+1.15) substrates. The value of $\Delta$H calculated for s(\%)=0.0 is $\approx$0.51 eV, and decreases about 23\%, down to $\approx$0.39 eV, for a tensile stress of s(\%)=+1.15. This behavior is very similar for the films synthesized under compressive strain, as shown in \ref{fig:4} b), with a progressive reduction of  $\Delta$H about 33\% for every 1\% of either positive or negative epitaxial stress. Hence both compressive and tensile epitaxial stress have a very similar effect on the reduction formation enthalpy of oxygen vacancies in electron-doped STO thin films, suggesting a common mechanism for the reduction of $\Delta$H under compressive and tensile strain. This is in agreement with ab-initio calculations by Choi et al.,\cite{ChoiNL} which suggested to a decrease of the band-gap energy under epitaxial strain (either positive or negative) as the main reason behind this effect.\cite{NeatonPRL} Indeed, creating an oxygen vacancy involves a redox process, as explained in equation (\ref{eq2}), regardless of whether the stress on the film is positive or negative.

\begin{figure}[h]
\begin{center}
\includegraphics[width=200pt]{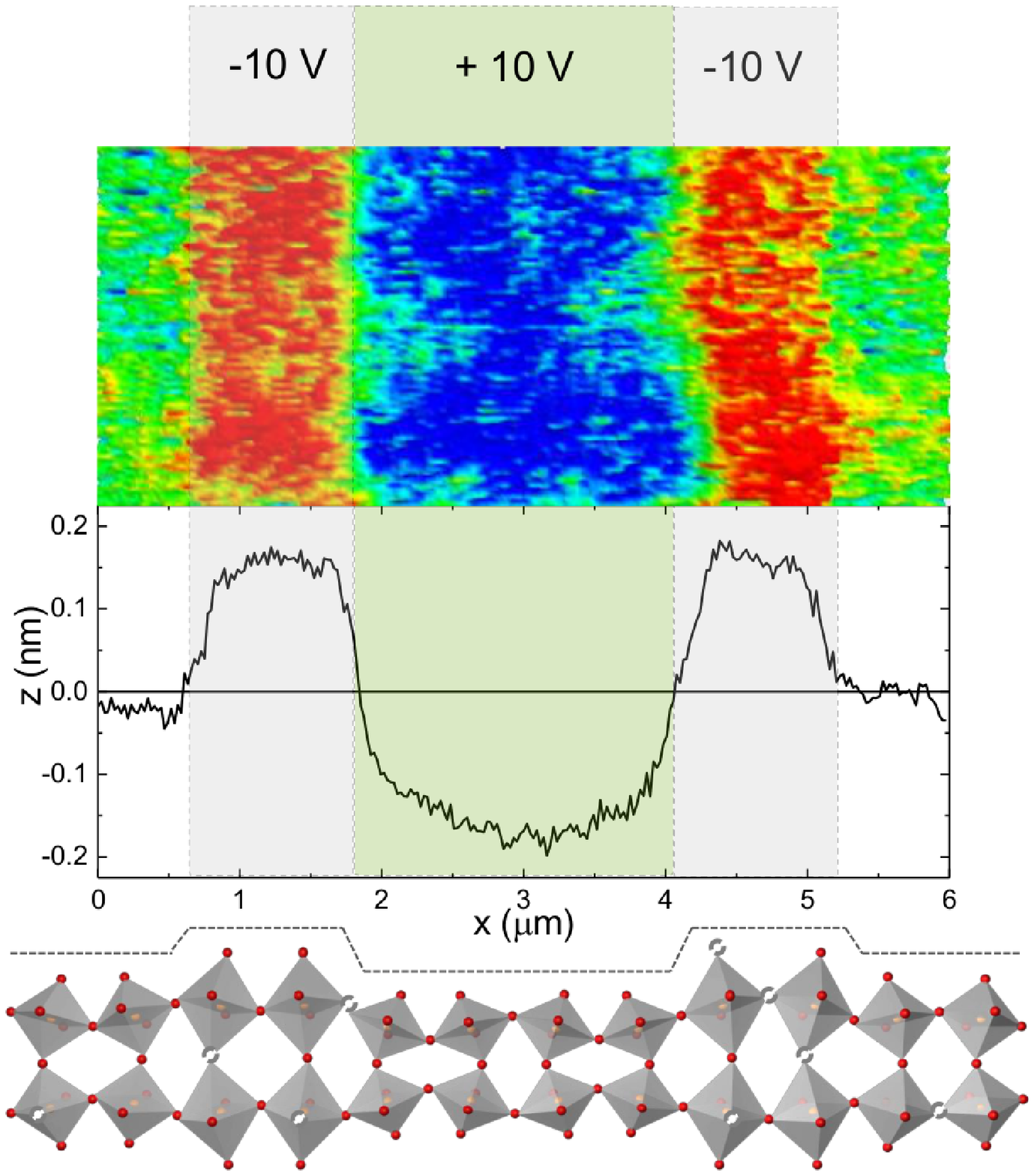}
\caption{AFM Topography image (top) and corresponding height profile (bottom) after aplying  +/- 10V on different regions of the sample, as indicated. The cartoon at the bottom represents the elongation (contraction) due to the accumulation (depletion) of oxygen vacancies (grey open circles).}\label{fig:5}
\end{center}
\end{figure}

Finally, once the parameters which control the formation of oxygen vacancies under different degrees of strain were established, we have studied the migration of vacancies throughout the film under the influence of an electric field. The change in the local concentration of vacancies was studied by EFM. Applying a positive/negative voltage to the sample is able to repeal or attract the positively charged oxygen vacancies, therefore creating regions of vacancy depletion/accumulation.

The results for a Nb:STO film deposited on DSO are shown in Figure \ref{fig:5}. Applying a local voltage of -10 V attracts the positively charged V$_O$, changing the electrostatic potential (see supporting information Figure S3). Although a similar effect over the local surface potential and conductivity was recently reported by Waser et Al., \cite{AndraNanoscale} we observed a clear change in the local volume of the sample, consistent with the local chemical expansion associated to the oxygen vacancies. This is further appreciated through the reversibility of the mechanical effect: application of a negative (positive) voltage produces a local expansion (contraction) of the sample, from its original value. The observed increase of the height in different samples ranges between $\approx$0.15-0.4 nm, roughly the size of a half to one unit cell size.
 Given the low diffusion coefficient of oxygen vacancies at room temperature in these oxides, they undergo an extremely low relaxation after removing the electric field. Therefore, these effects are stable for hours after removing the electric field.

The observation of this voltage-induced mechanical effect opens new possibilities to control the local concentration of oxygen vacancies (and hence the local conductivity) now by purely mechanical methods, i.e. mechanically assisted resistive switching.\cite{CBENanolett}

Further experiments to determine the kinetics of this effect, as well as the influence of surface termination (effectiveness of the electric field through SrO terminated with respect to TiO$_2$ terminated films\cite{Sitaputra}) are underway.


In summary, we have experimentally demonstrated the reduction of the formation energy of oxygen vacancies in films of SrTiO$_3$ under epitaxial stress. The similar trend observed under tensile and compressive stress calls for a common mechanism, probably a reduction in the band gap. We have also demonstrated that the pattern of TiO$_6$ rotations are mostly determined by the concentration of cationic vacancies unintentionally introduced during growth. This effect must be taken into account for a proper understanding of the changes in the electronic band-structure of this material under strain.
Oxygen vacancy accumulation/depletion can be achieved by an applied bias voltage, resulting in non-volatile changes of the local electrical potential and, more importantly, a change in the local volume. The latter opens the possibility of a mechanical control of the local concentration of oxygen vacancies, and associated properties, e.g. conductivity, catalytic properties, etc.


\begin{acknowledgments}
This work was supported the Ministry of Science of Spain (Project No. MAT2016-80762-R), the Conselleria de Cultura, Educacion e Ordenacion Universitaria (ED431F 2016/008, and Centro singular de investigacion de Galicia accreditation 2016-2019, ED431G/09) and the European Regional Development Fund (ERDF).L. I. B. also acknowledges the Ministry of Science of Spain for an FPI grant.
\end{acknowledgments}

\section{Supporting information}

\begin{table*}[t]
\centering
\caption{List of substrates used in this work for deposition of Nb-doped SrTiO$_3$ films under different degrees of epitaxial strain. Where 3.905 (\AA) and a$_s$ refer to the lattice parameter of bulk STO and to the pseudocubic lattice parameter of the substrate, respectively.}
\label{my-label}
\begin{tabular}{|l|l|l|}
\hline
       & \begin{tabular}[c]{@{}l@{}}Pseudocubic lattice parameter\\ (\AA)\end{tabular} & \begin{tabular}[c]{@{}l@{}}Strain \\ (\%=$\frac{a_s-3.905}{3.905}$$\times$100)\end{tabular} \\ \hline
(100) LaAlO$_3$ (LAO) & 3.791                                                                       & -2.91                                                                        \\ \hline
(100) (NdAlO$_3$)$_{0.39}$-(SrAl$_{0.5}$Ta$_{0.5}$O$_{3}$)$_{0.61}$(NSAT)
   & 3.840                                                                         & -1.66                                                                        \\ \hline
(100) (LaAlO$_3$)$_{0.29}$-(SrAl$_{0.5}$Ta$_{0.5}$O$_{3}$)$_{0.71}$(LSAT)   & 3.868                                                                       & -0.95                                                                        \\ \hline
(110) LaGaO$_3$ (LGO) & 3.885                                                                         & -0.51                                                                        \\ \hline
(100) SrTiO$_3$ (STO) & 3.905                                                                      & 0                                                                            \\ \hline
(110) DyScO$_3$ (DSO) & 3.950                                                                         & +1.15                                                                        \\ \hline
(110) GdScO$_3$ (GSO) & 3.970                                                                         & +1.66                                                                        \\ \hline
\end{tabular}
\end{table*}


\begin{figure}[h!]
\begin{center}
\includegraphics[width=200pt]{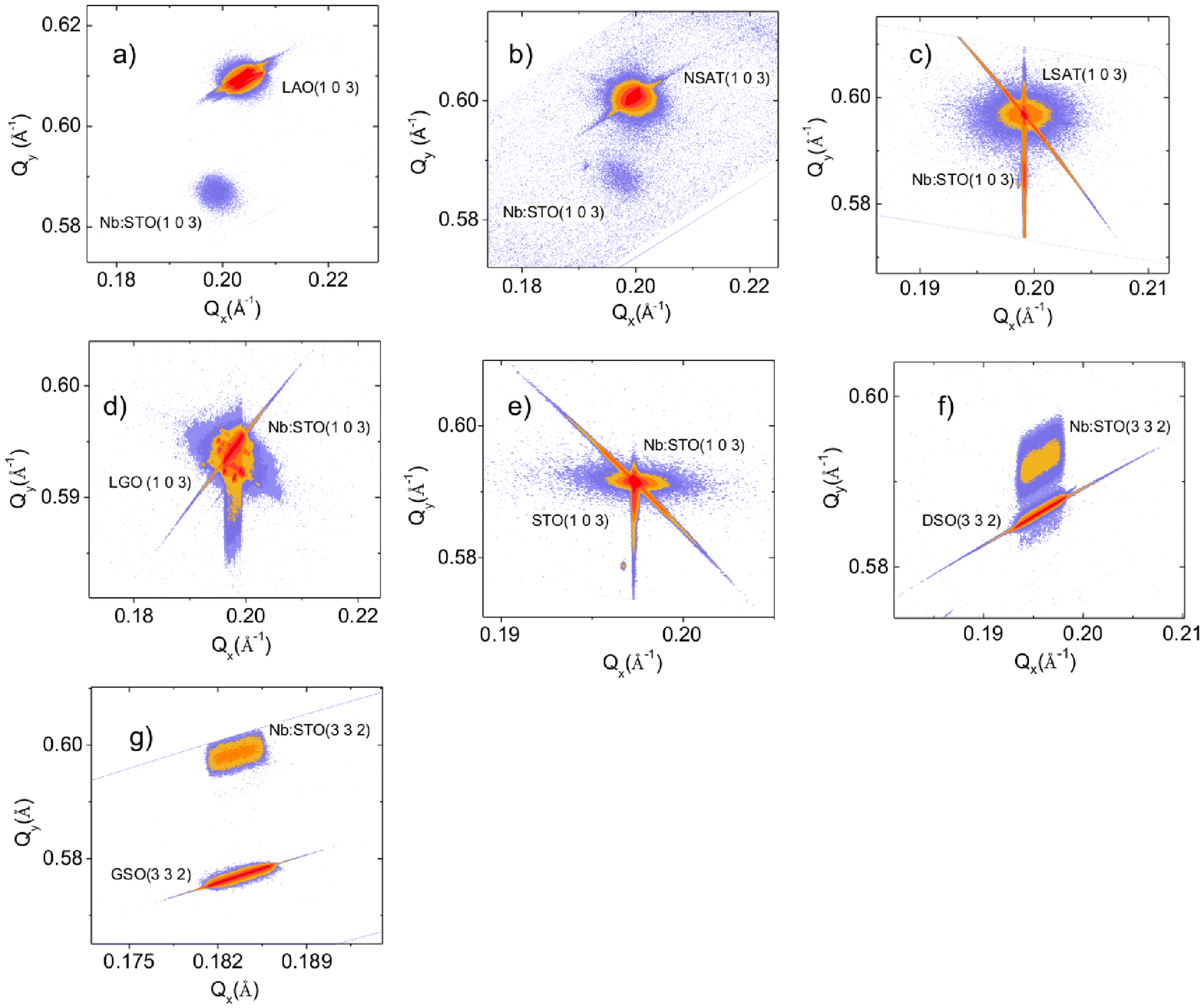}
\caption{High-resolution X-ray Reciprocal Space Maps (RSM) around different Bragg peaks for Nb-STO thin-films deposited in the different substrates used in this work.}
\end{center}
\end{figure}


\begin{figure}[h!]
\begin{center}
\includegraphics[width=200pt]{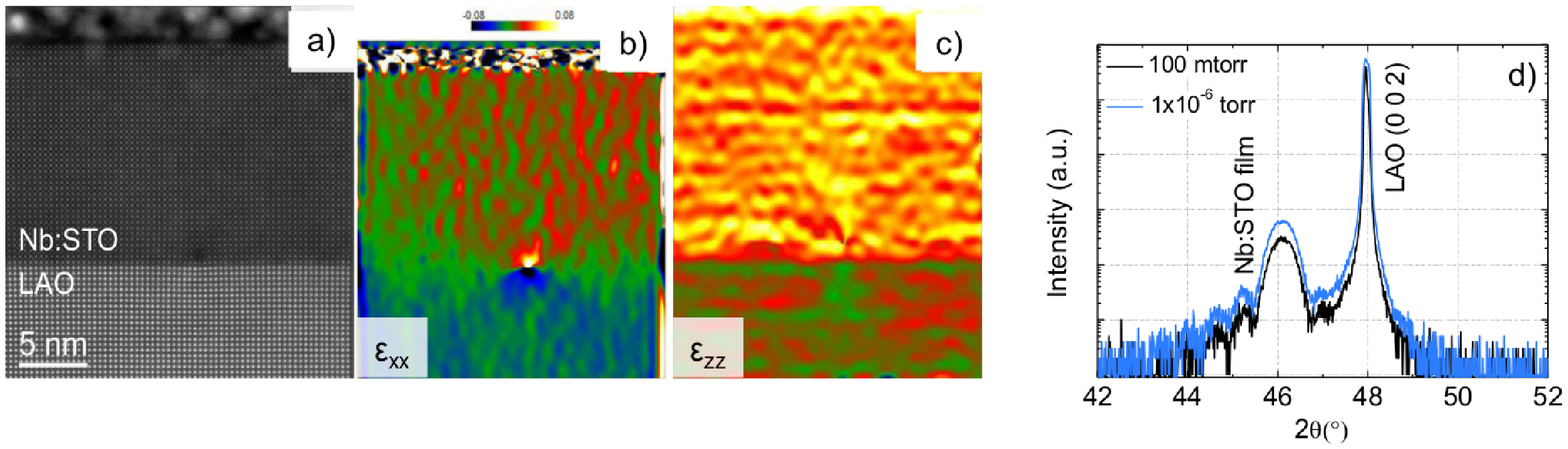}
\caption{HAADF-STEM image of as-grown Nb:STO thin films deposited on LAO (a). The corresponding GPA analysis showing the in-plane $\varepsilon_{xx}$ and out of plane $\varepsilon_{zz}$ elongation with respect to the substrate are shown in b) and c). d)  XRD $\theta{}$-2$\theta{}$ scan of the sample as-grown at 100 mTorr and after the annealing process at 1x10$^{-6}$ Torr, showing the negligible effect on the lattice parameters.}
\end{center}
\end{figure}


\begin{figure}[h!]
\begin{center}
\includegraphics[width=200pt]{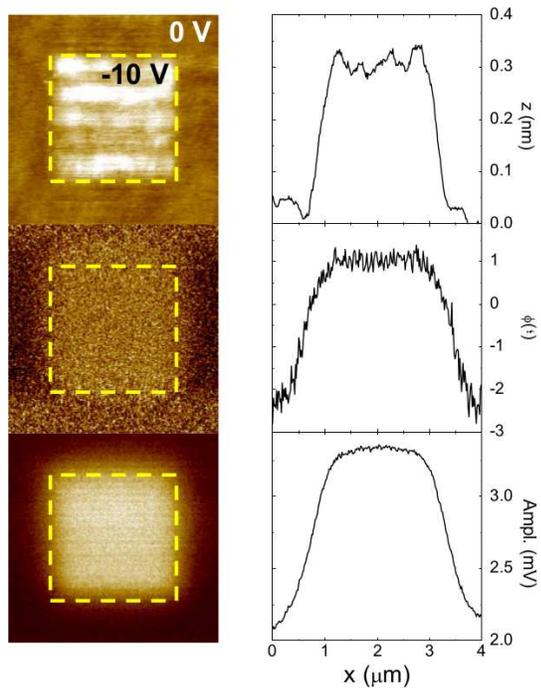}
\caption{Topography (top), phase (middle) and amplitude (bottom), along with their respective profiles from an Electrostatic Force Microscopy (EFM) experiment on Nb:STO deposited on DyScO$_3$. The applied voltage was - 10 V inside the area delimited by the yellow dashed line.}
\end{center}
\end{figure}

\newpage
%
\end{document}